\documentclass[longbibliography,a4paper,floatfix,noshowpacs,jcp,reprint,preprintnumbers,superscriptaddress]{revtex4-1}
\pdfoutput=1

\usepackage[utf8]{inputenc}
\usepackage[english]{babel}
\usepackage{amsfonts,amsmath,amssymb}
\usepackage{graphicx}
\usepackage{bm} 
\usepackage{mathrsfs}
\usepackage{subfigure}
\usepackage{textcomp}
\usepackage{hyperref}
\usepackage[flushleft]{threeparttable}
\usepackage[per-mode=symbol]{siunitx}
\usepackage[version=3]{mhchem}
\graphicspath{{../figures/}}
\usepackage{upgreek}

\DeclareSIUnit\intensity{\watt\per\centi\meter\squared}
\DeclareSIUnit\fieldstrength{\volt\per\centi\meter}

\newcommand{\degree}{\ensuremath{^\circ}}%

\newlength{\figwidth}
\newlength{\figwidthsmall}
\let\orgautoref\autoref
\providecommand{\Autoref}{%
  \def\equationautorefname{Equation}%
  \def\figureautorefname{Figure}%
  \def\subfigureautorefname{Figure}%
  \def\sectionautorefname{Section}%
  \orgautoref}
\renewcommand{\autoref}{%
  \def\equationautorefname{Eq.}%
  \def\figureautorefname{Fig.}%
  \def\subfigureautorefname{Fig.}%
  \def\sectionautorefname{Sec.}%
  \orgautoref}

\usepackage{color}
\usepackage[normalem]{ulem}
\definecolor{darkgreen}{rgb}{0.0,0.7,0.0}

\begin{document}

\title{Femtosecond Laser Induced Coulomb Explosion Imaging of Aligned OCS Oligomers inside Helium Nanodroplets}
\author{James D.\ Pickering}
\author{Benjamin Shepperson}
\author{Lars Christiansen}
\author{Henrik Stapelfeldt}
\email{henriks@chem.au.dk}
\affiliation{Department of Chemistry, Aarhus University, Langelandsgade 140, 8000 Aarhus C}
\date{\today}

\begin{abstract}

	Dimers and trimers of carbonyl sulfide (OCS) molecules embedded in helium nanodroplets are aligned by a linearly polarized \SI{160}{\pico\second} long moderately intense laser pulse and Coulomb exploded with an intense \SI{40}{\femto\second} long probe pulse in order to determine their structures. For the dimer, recording of 2D images of \ce{OCS^+} and \ce{S^+} ions and covariance analysis of the emission directions of the ions allow us to conclude that the structure is a slipped-parallel shape similar to the structure found for gas phase dimers. For the trimer, the \ce{OCS^+} ion images and corresponding covariance maps reveal the presence of a barrel-shaped structure (as in gas phase) but also other structures not present in the gas phase, most notably a linear chain structure.

\end{abstract}
\maketitle

\section{Introduction}\label{sec:introduction}

Noncovalently bonded complexes of molecules are important for a number of physical and chemical phenomena including the shape of bioactive molecules~\cite{becucci_high-resolution_2016}, chiral recognition~\cite{zehnacker_chirality_2008}, exciplex formation~\cite{yip_excimer/exciplex_1996}, and bimolecular reactions~\cite{wheeler_probing_2000}. Experimentally, noncovalently bonded molecular complexes have been studied in the gas phase using molecular beams from supersonic expansions. Here the typical temperature of a few Kelvin and the collision-free conditions ensure that the weakly bonded complexes remain intact on the time scale of experiments. Alternatively, molecular complexes can be formed inside helium nanodroplets where the low (\SI{380}{\milli\kelvin}) temperature, and the possibility to add molecules to the droplet sequentially, enables formation of a wide range of complexes normally inaccessible using supersonic gas expansion methods~\cite{choi_infrared_2006,yang_helium_2012,bellina_proton_2015,douberly_the_2005}.

Prior studies of molecular complexes in the gas phase and in \ce{He} droplets have primarily been undertaken using high-resolution frequency-resolved techniques~\cite{nesbitt_high-resolution_1988,moazzen-ahmadi_spectroscopy_2013,nauta_infrared_2001,moore_binary_2001,choi_multiple_2005,raston_infrared_2017,sadoon_infrared_2016,sulaiman_infrared_2017}. These techniques have provided a deep insight into the structure of such complexes, but the high frequency resolution necessitates low temporal resolution. This is a drawback if studies of molecular dynamics are desired, as to follow reaction dynamics in real-time requires a probe with sub-picosecond temporal resolution. A complementary technique to such frequency resolved methods is laser-induced Coulomb explosion~\cite{yatsuhashi_multiple_2018}. Here an intense femtosecond (fs) laser pulse multiply ionizes a molecule or a molecular complex breaking it into cationic fragments. If the fragment ions recoil along the original bond axes of their parent molecule (axial recoil approximation) the molecular structure can often be identified by recording of the fragment ion velocity vectors, and analysing correlations between them.

In the gas phase, laser-induced Coulomb explosion has been used to accurately determine the structure of van der Waals bonded dimers consisting of linear molecules, such as \ce{N_2}, and noble gas atoms such as \ce{Ar}~\cite{wu_structures_2012,wu_communication:_2014}. Recently, our group demonstrated that the structure of the \ce{CS_2} dimer formed inside He nanodroplets could be determined by Coulomb explosion~\cite{pickering_alignment_2018}. In that study laser-induced alignment~\cite{stapelfeldt_colloquium:_2003} was used prior to the explosion process to fix the dimers in well-defined positions with respect to the imaging detector recording the velocity of the fragment ions. This proved highly advantageous for the structural determination. The purpose of the present work is to extend the first studies on the \ce{CS_2}  dimers. It is done by applying the alignment and Coulomb explosion techniques to dimers and trimers of \ce{OCS} in \ce{He} nanodroplets.

\section{Experimental and Analysis Methods}

\subsection{Experimental Setup}\label{sec:methods}
\begin{figure}
\centering
\includegraphics[width=\columnwidth]{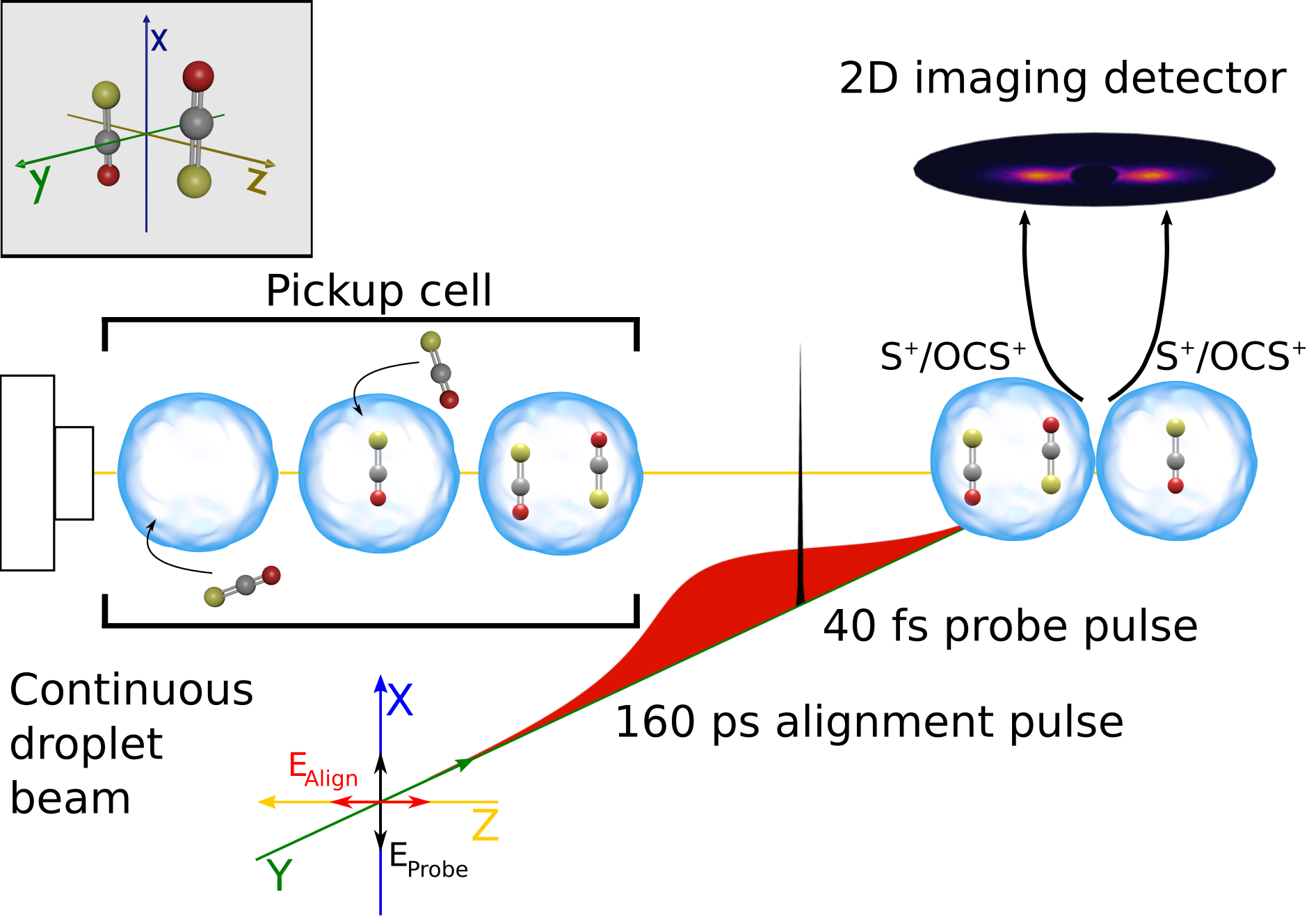}
\caption{Schematic depicting key elements of the experiment. Here the alignment (probe) pulse is depicted polarized along the $Z$ ($X$) axis. The electrodes in the VMI spectrometer projecting the ions onto the detector are not shown. The gray insert in the top-left shows the molecular frame, and the gas-phase \ce{OCS} dimer structure.}\label{fig:experimental_setup}
\end{figure}

The experimental setup has been described in detail before~\cite{shepperson_strongly_2017}, and only important aspects will be mentioned here. \Autoref{fig:experimental_setup} shows a schematic of the setup used. A beam of \ce{He} droplets is formed by continuously expanding \ce{He} gas at \SI{14}{\kelvin} and \SI{30}{\bar} backing pressure into vacuum through a \SI{5}{\micro\metre} nozzle. The droplets consist of 10000 \ce{He} atoms on average~\cite{toennies_superfluid_2004}, and pass through a pickup cell containing \ce{OCS} gas. The partial pressure of \ce{OCS} in this cell is controlled using a high-precision leak valve. This affords rudimentary control of the clustering inside the droplets, discussed further in \autoref{sec:doping}. The doped droplet beam then enters the interaction region where it is crossed at 90\degree\,by two focussed laser beams originating from the same Ti-Sapphire laser system. The uncompressed output of this laser system is used to provide \SI{160}{\pico\second} (FWHM) pulses ($\lambda_\text{center} = \SI{800}{\nano\metre}, I_0 = \SI{8E11}{\intensity}$) to align the \ce{OCS} oligomers. The compressed output of this laser system is used to provide \SI{40}{\femto\second} (FWHM) probe pulses ($\lambda_\text{center} = \SI{800}{\nano\metre}, I_0 = \SI{3E14}{\intensity}$) to Coulomb explode the \ce{OCS} oligomers. The nascent fragment ions are subsequently focussed in a velocity-map imaging (VMI) spectrometer onto a 2D imaging detector. The detector consists of two microchannel plates backed by a P47 phosphor screen, the light from which is detected by a CCD camera. The experiments are run at a repetition rate of \SI{1}{\kilo\hertz}, dictated by the repetition rate of the laser system. Throughout the following discussion, we refer to the molecule-fixed frame as the $(x,y,z)$ frame, and to the space-fixed (laboratory) frame as the $(X,Y,Z)$ frame (see \autoref{fig:experimental_setup}).

\subsection{Covariance Analysis}\label{sec:covariance}
The attainable information from ion imaging data can be greatly augmented by identifying correlations between the velocities or emission angles of the ions produced. One approach to this is covariance analysis. The covariance between two ion observables $X$ and $Y$, $cov(XY)$, is defined as:

\begin{equation}\label{eq:covar}
cov(XY) = \langle XY \rangle - \langle X \rangle \langle Y \rangle
\end{equation}

where $\langle XY \rangle$ is the expectation value of observing the two observables $X$ and $Y$ together; and $\langle X \rangle$, $\langle Y \rangle$ are the expectation values of each of the two observables, observed separately. Applying \autoref{eq:covar} to the emission angles of recoiling ion fragments allows correlations between the emission angles of ion fragments to be calculated. The resulting covariance data is presented as an angular covariance map, where the magnitude of the covariance between two ions recoiling at angles $\theta_1$ and $\theta_2$ is plotted on a grid~\cite{frasinski_covariance_1989,hansen_control_2012,slater_covariance_2014}. A large value of the covariance between two ion emission angles shows that ion fragments are more commonly ejected at these angles than others. For example, a covariance signal at angles $\theta_1 = \theta_2 + 180$\degree\,would indicate two ion fragments recoiling in opposite directions to each other. This technique allows information that is otherwise hidden in the raw data to be revealed, and greatly enhances the utility of ion imaging as a structural probe~\cite{christensen_dynamic_2014,burt_communication_2018}.

\section{Results and Discussion}\label{sec:results}

\subsection{Doping Regimes}\label{sec:doping}

\begin{figure}
\centering
\includegraphics[width=\columnwidth]{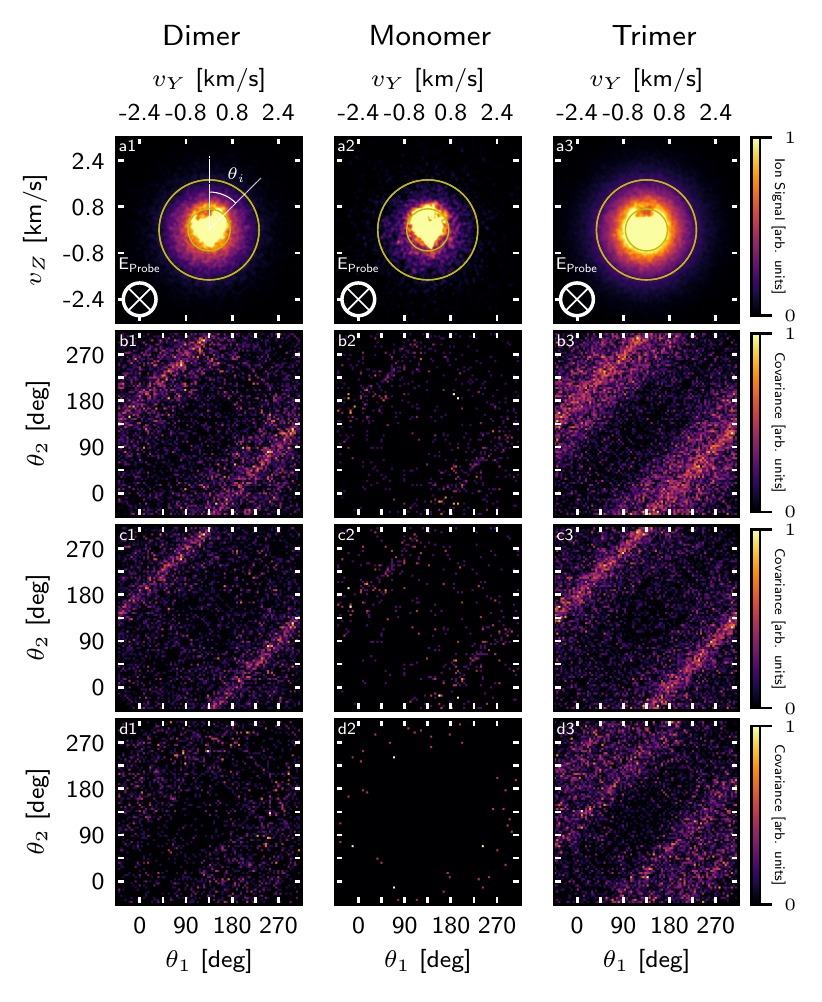}
\caption{(a1)-(a3) \ce{OCS^+} ion images and (b1)-(d3) corresponding angular covariance maps. The polarization state of the probe pulse is given in the lower left corner of each image. Data in column (1) [(2)] \{(3)\} were recorded under the dimer [monomer] \{trimer\} doping condition (see text). Angular covariance maps in row (b) [(c)] \{(d)\} were calculated using all [dimer] \{trimer\} radii (see text). $v_Y$ and $v_Z$ are the velocity components of the speed $v$ along the $Y$ and $Z$ axes respectively.}\label{fig:doping_conditions}
\end{figure}

\Autoref{fig:doping_conditions}\,(a1) shows an \ce{OCS^+} ion image recorded using a linearly polarized probe pulse polarized along the $X$ axis (perpendicular to the detector plane). An intense signal is present in the central portion of the image, and some weaker signal occurrs at larger radii. The \ce{OCS^+} ions produced via ionization of \ce{OCS} molecules in droplets containing only single \ce{OCS} molecules will necessarily have very low speed. Thus, we ascribe the intense central signal as resulting from ionization of \ce{OCS} monomers\,\footnote{There is also possibly a small signal from \ce{He_{15}+}, which will overlap the \ce{OCS+} in this central region}. However, \ce{OCS^+} ions arising from Coulomb explosion of oligomers of \ce{OCS} will have a non-negligible speed due to the Coulombic repulsion between two or more \ce{OCS^+} ions in the He droplet (see \autoref{sec:ker}). We therefore ascribe the signal occurring at radii outside the innermost yellow circle in \autoref{fig:doping_conditions}\,(a1) ($v >$ \SI{0.6}{\kilo\metre\per\second}, where speed $v = \sqrt{(v_Y^2 + v_Z^2)}$\,\footnote{Calibration of the spectrometer allows the speed $v$ (\si{\metre\per\second}) of an ion hitting a pixel of radius $r$ (pixels) to be calculated as $v = ar$, where $a = 15.17$ is a scaling factor dependent upon the mass of the ion fragment and the voltages applied to the ion optics.}) as arising from Coulomb explosion of \ce{OCS} oligomers.

\Autoref{fig:doping_conditions}\,(b1) shows the angular covariance map calculated over all ion hits outside the innermost yellow circle in \autoref{fig:doping_conditions}\,(a1). This shows that there are clear correlations between \ce{OCS+} ions recoiling at $\theta_1 = \theta_2 + 180$\degree\,(where $\theta_i$ is the angle an ion $i$ hits on the detector, relative to upwards vertical, illustrated in \autoref{fig:doping_conditions}\,(a1)). This is consistent with a two-body breakup, and therefore Coulomb explosion of an \ce{OCS} dimer (\ce{(OCS)_2}) into a pair of \ce{OCS^+} ions. One can also see some faint signal at different angles, at $\theta_1 = \theta_2 + 120$\degree\,and $\theta_1 = \theta_2 + 240$\degree. This is inconsistent with a two-body break up as expected for \ce{(OCS)_2}, but would be consistent with three-body break up of an \ce{OCS} trimer (\ce{(OCS)_3}), thus we tentatively ascribe this signal as such. \Autoref{fig:doping_conditions}\,(c1) shows the angular covariance map now calculated over all ion hits between the innermost and outermost (\SI{0.6}{\kilo\metre\per\second} $< v <$ \SI{1.4}{\kilo\metre\per\second}) yellow circles in \autoref{fig:doping_conditions}\,(a1). Here there is a marked reduction in the number of ions seen at angles apart from $\theta_1 = \theta_2 + 180$\degree. Similarly, \autoref{fig:doping_conditions}\,(d1) shows the angular covariance map calculated over all ion hits outside of the outermost yellow circle in (a1) ($v >$\SI{1.4}{\kilo\metre\per\second}). Here none of the $\theta_1 = \theta_2 + 180$\degree~ signal is present, instead only some faint correlations consistent with the aforementioned breakup of \ce{(OCS)_3} present.

We conclude that selecting the radii in this way allows the ions resulting from Coulomb explosion of \ce{(OCS)_2} to be distinguished from those originating from explosion of \ce{(OCS)_3}, as an \ce{OCS^+} ion arising from \ce{(OCS)_3} will experience Coulombic repulsion from two other \ce{OCS^+} partners, resulting in a higher speed. In contrast, an \ce{OCS^+} ion arising from Coulomb explosion of \ce{(OCS)_2} will only experience repulsion from one other \ce{OCS^+} partner, resulting in a lower speed in the detector plane. This can be quantified with electrostatic arguments, outlined in \autoref{sec:ker}. On this basis, we define the radii used in (b1) as `all radii' (outside the innermost circle, $v >$ \SI{0.6}{\kilo\metre\per\second}); radii used in (c1) (between the innermost and outermost circles, \SI{0.6}{\kilo\metre\per\second} $< v <$ \SI{1.4}{\kilo\metre\per\second}) as the `dimer radii'; and radii used in (d1) (outside of the outermost circle, $v >$ \SI{1.4}{\kilo\metre\per\second}) as the `trimer radii'.

It is clear that the majority of the signal in \autoref{fig:doping_conditions}\,(a1) arises from \ce{OCS} monomers, but that there are a substantial amount of dimers present and a smaller amount of trimers. We define this doping-pressure region as the `dimer-doping-condition', in line with previous work~\cite{pickering_alignment_2018}. One then expects that it would be possible to enhance or reduce various parts of this signal by altering the pressure of the \ce{OCS} gas in the pickup cell. \Autoref{fig:doping_conditions}\,(a2) shows an \ce{OCS^+} image recorded with a much lower partial pressure of \ce{OCS} in the doping cell compared to the case of \autoref{fig:doping_conditions}\,(a1). It is evident from the angular covariance maps (\autoref{fig:doping_conditions}\,(b2)-(d2)) that there is markedly less \ce{(OCS)_2} present here, and very little \ce{(OCS)_3}. This is expected, as the number of dimers and trimers will be reduced as the pressure in the doping cell is reduced~\cite{toennies_superfluid_2004}. There is still a visible monomer contribution in \autoref{fig:doping_conditions}\,(a2), thus we define this region as the `monomer-doping-condition'. Finally, for the \ce{OCS^+} image in \autoref{fig:doping_conditions}\,(a3) the \ce{OCS} pressure in the pickup cell was increased compared to the case of \autoref{fig:doping_conditions}\,(a1). Now it is clear that there are more ions outside the outermost yellow circle, and the covariance maps \autoref{fig:doping_conditions}\,(b3)-(d3) show an enhanced dimer and trimer signal compared to covariance maps \autoref{fig:doping_conditions}\,(b1)-(d1). As there are clearly substantially more trimers present under this doping condition, we define this as the 'trimer-doping-condition'.

\Autoref{tab:counts} gives the total number of ions inside the innermost yellow circle (monomers); between the two yellow circles (dimers); and outside the outermost yellow circle (trimers). The latter two of these radii correspond to the aforementioned dimer radii and trimer radii, where the former condition (inside the innermost circle) is referred to as the `monomer radii'. Under all doping conditions there are more monomers than dimers, and more dimers than trimers, but it is possible to increase the relative proportions of dimers and trimers by increasing the doping pressure.

It is pertinent to note two things. Firstly, the doping of molecules into He droplets is known to be governed by Poisson statistics~\cite{toennies_superfluid_2004,yang_helium_2012} thus it is never possible to completely isolate one oligomer of \ce{OCS} for study through altering the doping pressure alone. Secondly, due to the statistical nature of the pickup process, it is possible that contributions from higher oligomers are present when in doping regimes with substantial amounts of trimers. However, our imaging detector is substantially less sensitive in the central area (monomer radii) than in the outer regions (by an estimated factor of 2 or 3) due to the consistently high ion flux incident on this region over the lifetime on the detector. This is expected to lead to a considerable underestimate of the amount of \ce{OCS} monomers present. Consequently, we believe we are in a regime where contributions from higher oligomers are minimised as far as possible.

\begin{center}
\begin{table}
	\begin{ruledtabular}\begin{tabular}{cccc}
Doping Condition & Monomer (\%) & Dimer (\%) & Trimer (\%) \\ \hline
Monomer & 66.98 & 26.65 & 6.37 \\
Dimer & 54.88 & 34.96 & 10.16 \\
Trimer & 44.36 & 39.89 & 15.76 \\
	\end{tabular}\end{ruledtabular}
\caption{The relative amounts of \ce{OCS^+} ions hitting the detector at different radii at the different specified doping conditions. The proportions are given relative to the total number of \ce{OCS^+} ions hitting the detector.}\label{tab:counts}
\end{table}
\end{center}

\subsection{Quantitative Analysis of Kinetic Energy Release}\label{sec:ker}
\begin{figure}
\centering
\includegraphics[width=\columnwidth]{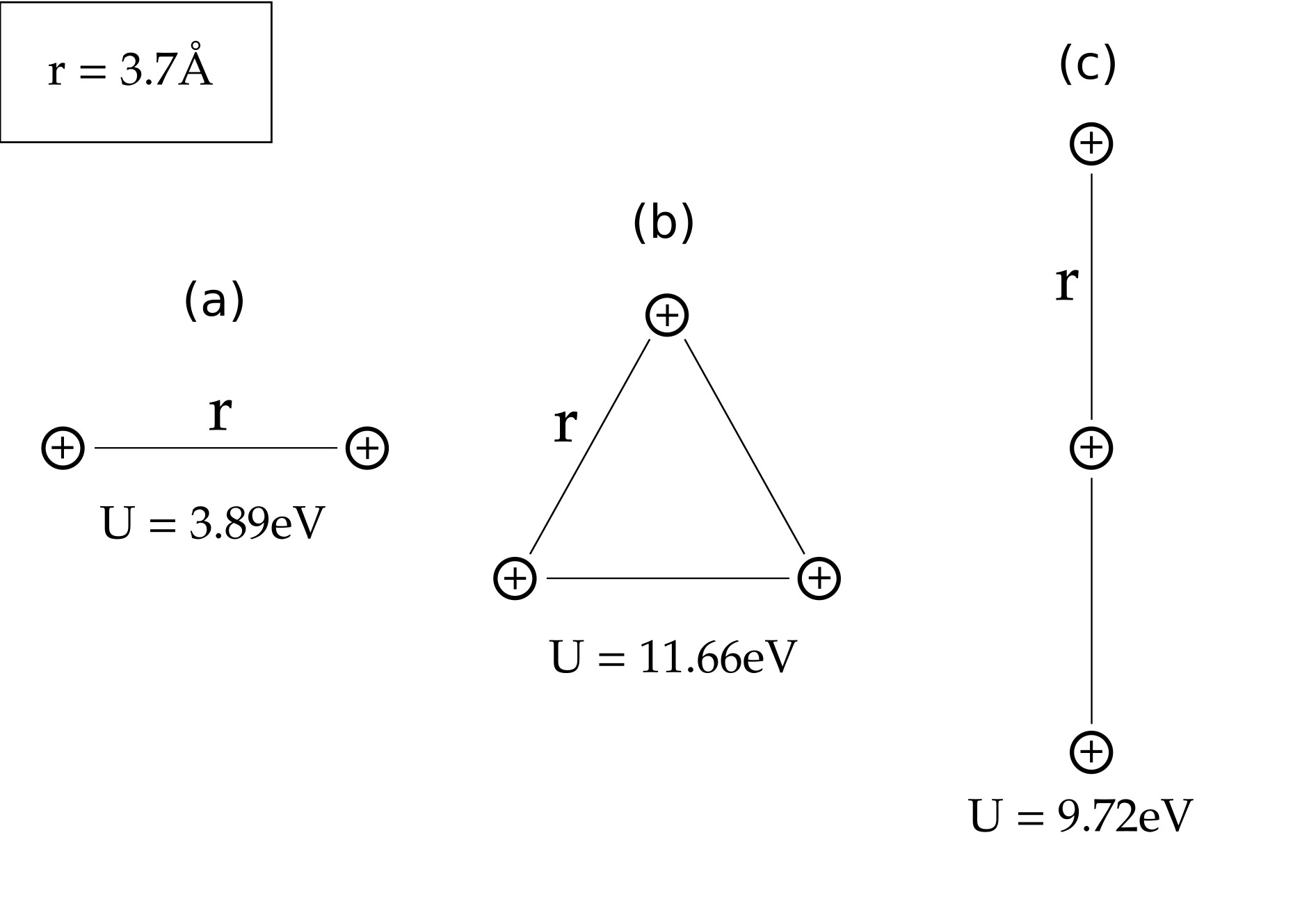}
\caption{Sketches of arrangements of point charges in the dimer (a); triangular trimer (b); and linear trimer (c) geometries, used to estimate the potential energy $U$ stored. The total potential energy $U$ stored in each geometry is given below each drawing. The separation between two adjacent point charges $r$ was taken to be \SI{3.7}{\angstrom} in all cases.}\label{fig:charged_structures}
\end{figure}

The assignment of monomer, dimer, and trimer radii can be justified using electrostatic arguments. Assuming all \ce{OCS} molecules in an oligomer are singly ionized, and that each \ce{OCS^+} ion can be modelled as a point charge, the potential energy $U(N)$ stored in an oligomer consisting of $N$ \ce{OCS^+} ions, at the moment of ionization, is given by:

\begin{equation}\label{eq:coulomb}
	U(N) = \frac{\mathrm{e}^2}{8\pi\upepsilon_0} \sum\limits_{i=1}^N \sum\limits_{j=1}^{N(j \neq i)} \frac{1}{r_{ij}}
\end{equation}

where $\mathrm{e}$ is the elementary charge, $r_{ij}$ is the distance between ions $i$ and $j$, and $\upepsilon_0$ is the vacuum permittivity. Assuming that the potential energy $U$ is completely converted to kinetic energy via Coulombic repulsion, we can then estimate the kinetic energy release that will be observed by considering different arrangements of point charges. \Autoref{fig:charged_structures} shows possible arrangements of point charges that are based on the observed structures of \ce{(OCS)_2} and \ce{(OCS)_3} in the gas phase (images (a) and (b) respectively), and an arrangement based on a linear structure of \ce{(OCS)_3} for comparative purposes (image (c)). The total potential energy stored in each structure is annotated below each structure, where it is clear that the triangular structure (image (a)) stores the most electrostatic potential energy, and the dimer structure (image (b)) the least. The separation between each point charge was taken to be \SI{3.7}{\angstrom}, a reasonable approximation to the true separation in the gas-phase~\cite{bone_an_1993}, as the true separation in \ce{He} droplets has not been determined.

We now consider how the potential energy will be distributed into each constituent charge when they undergo Coulombic repulsion. In the triangular structure, the potential energy will be equally shared between each constituent ion, such that each \ce{OCS^+} ion would gain \SI{3.89}{\electronvolt} of kinetic energy (corresponding to a maximum speed of around \SI{3.5}{\kilo\metre\per\second} in the detector plane). Similarly, for the dimer structure each \ce{OCS^+} ion gains \SI{1.95}{\electronvolt} of kinetic energy (a maximum speed of around \SI{2.5}{\kilo\metre\per\second} in the detector plane). However, in the linear trimer structure, the central \ce{OCS^+} ion would experience an equal and opposite repulsion from each terminal \ce{OCS^+} ion, such that it would gain no net kinetic energy; thus the total stored potential is shared equally between each terminal \ce{OCS^+} ion, each gaining \SI{4.86}{\electronvolt} of kinetic energy (a maximum speed of around \SI{4}{\kilo\metre\per\second} in the detector plane). This corroborates the observation that it is possible to select ion events arising from explosion of specific oligomers based on their different speeds in the detector plane.

It is noteworthy that these calculated energies lead to speeds that are substantially greater than the empirical speeds discussed in \autoref{sec:doping}. For example, it was observed that ions with a speed above \SI{1.4}{\kilo\metre\per\second} resulted from the explosion of \ce{OCS} trimers. This speed is vastly lower than the calculated speed of \SI{3.5}{\kilo\metre\per\second}, and similar observations can be made for other radii. We attribute the bulk of this effect to the kinetic energy lost by the recoiling \ce{OCS^+} ions as they scatter off \ce{He} atoms on their way out of the droplet~\cite{braun_photodissociation_2007-1}. It is also possible that in \ce{He} droplets the separation between \ce{OCS} molecules within the different clusters differs slightly from that observed in the gas phase; however this effect is expected to be minor compared to the effect of the He scattering.

\subsection{\ce{OCS} Dimer Results}\label{sec:dimer}
\begin{figure}[h!]
\centering
\includegraphics[width=\columnwidth]{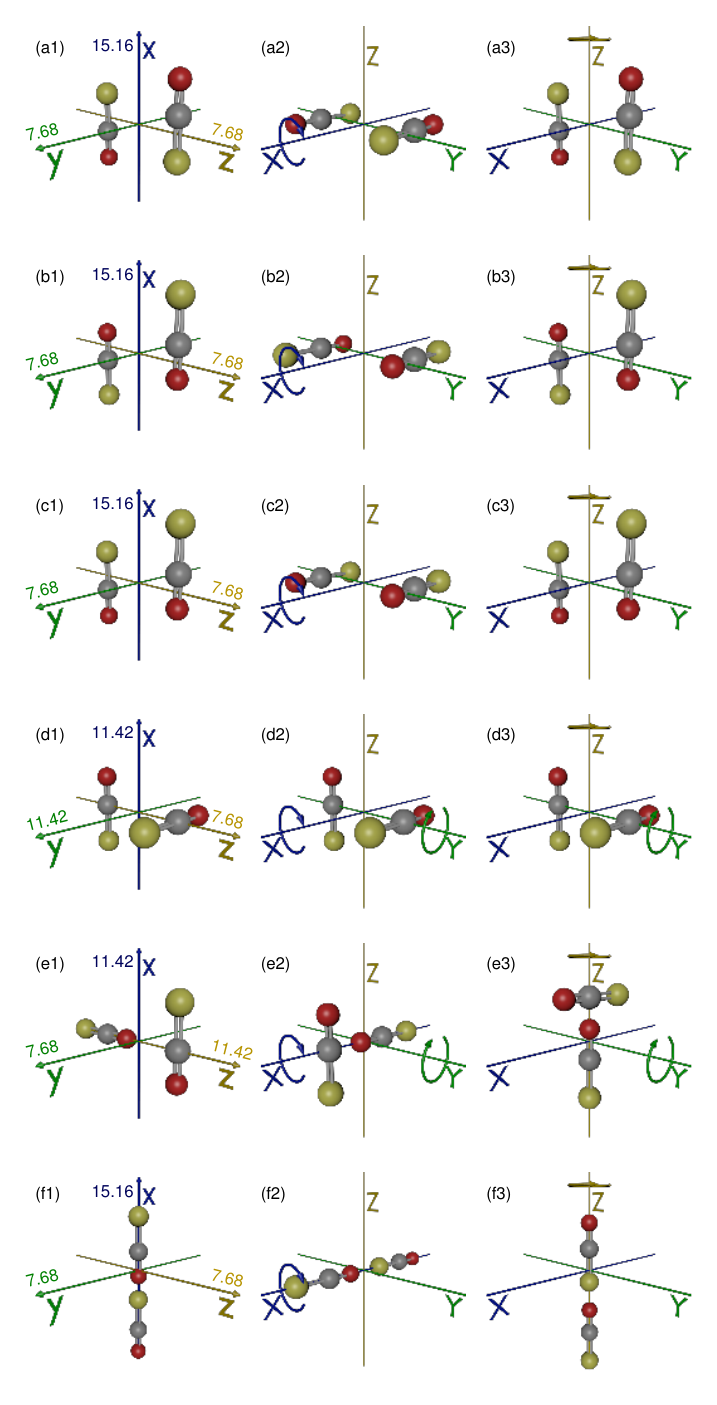}
\caption{(a1)-(f1) Sketches of potential \ce{OCS} dimer structures in the molecular ($x,y,z$) frame, with their polarizability elements ($\alpha_x$, $\alpha_y$, $\alpha_z$) annotated on the relevant axis, in units of $\AA^3$ (see text). (a2)-(f2) [(a3)-(f3)] Sketches of how each structure would align in the laboratory ($X,Y,Z$) frame with an aligning laser field polarized along the $X$ [$Z$] axis. The detector is placed in the $YZ$ plane. Curved arrows indicate axes about which rotation is possible.}\label{fig:potential_structures}
\end{figure}

We now show how our experimental data allow us to identify the structure of the \ce{OCS} dimer, inside \ce{He} droplets. Our strategy is to consider a number of potential candidates for the structures and investigate if they are consistent with our observations. The candidates, with their structures sketched in \autoref{fig:potential_structures}\,(a1)-(f1), were chosen due to their prior observation in the gas phase (a)-(c); or due to other small triatomics adopting analogous structures in the gas phase (d)-(f)~\cite{moazzen-ahmadi_spectroscopy_2013,nauta_nonequilibrium_1999-1,rezaei_spectroscopic_2011}.
The slipped-parallel sulfur-in structure shown in \autoref{fig:potential_structures}\,(a1) has been identified as the gas phase ground-state structure~\cite{moazzen-ahmadi_spectroscopy_2013}, but it is known that He droplets can anneal non-ground state structures~\cite{nauta_nonequilibrium_1999-1,nauta_formation_2000}. Thus, we consider all six possible structures initially.

To interpret the experimental data it is necessary to predict how each dimer structure considered aligns when exposed to the alignment pulse. This is possible if the polarizability tensor is known. Therefore, for each structure in \autoref{fig:potential_structures}\,(a1)-(f1), the polarizability components along each molecular axis ($\alpha_x$, $\alpha_y$, $\alpha_z$) are given. The polarizability components for each dimer structure were calculated by summing the polarizability components of the two monomers, with each rotated relative to each other as necessary. The neglect of the electronic interaction between the monomer units in each dimer will not qualitatively change the polarizability tensor of the dimer. The polarizability components of the \ce{OCS} monomer were taken to be $\alpha_{\parallel} = \SI{7.58}{\angstrom^3}, \alpha_{\perp} = \SI{3.84}{\angstrom^3}$~\cite{national_institute_of_standards_and_technology_computational_nodate}.

\Autoref{fig:potential_structures}\,(a2)-(f2) shows how each potential structure would align in the laboratory ($X,Y,Z$) frame, under an alignment field linearly polarized along the $X$ axis. A linearly polarized alignment field will align the most polarizable molecular axis (MPA) to the polarization direction~\cite{stapelfeldt_colloquium:_2003}. If there is not one single MPA, but two equally polarizable axes and one less polarizable axis, then the system will align such that either one of the most polarizable axes are parallel to the alignment field at any one time. To illustrate this, curved arrows on each sketch show axes where rotation (free or hindered) is possible under the alignment field. The detector is placed in the $YZ$ plane (see \autoref{fig:experimental_setup}). In cases where there are multiple possible aligned geometries, only one is shown in the sketch. \autoref{fig:potential_structures}\,(a3)-(f3) shows analogous sketches for an alignment field linearly polarized along the $Z$ axis.

\begin{figure}
\centering
\includegraphics[width=\columnwidth]{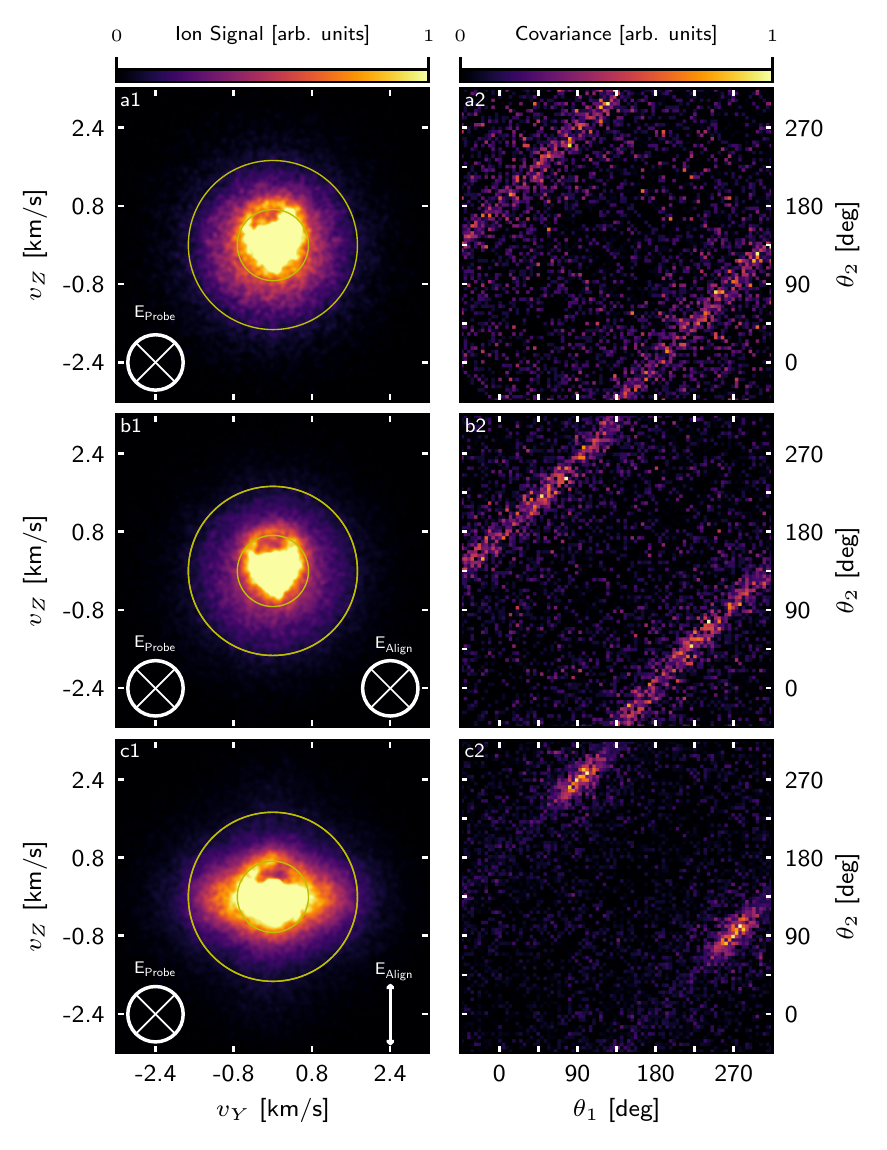}
\caption{(a1)-(c1) \ce{OCS^+} ion images and (a2)-(c2) corresponding angular covariance maps. All images (a1)-(c1) were taken under the dimer-doping-condition, and all covariance maps (a2)-(c2) were calculated using the dimer radii. The polarization state of the probe (alignment) laser is shown in the lower left (right) corner of the ion images.}\label{fig:dimer_structure}
\end{figure}

\Autoref{fig:dimer_structure}\,(a1) shows an \ce{OCS^+} ion image, recorded under the dimer-doping-condition, with the probe laser polarized along the $X$ axis, and no alignment laser present. There is clear circular symmetry in the image, but this image does not allow the dimer structure to be determined directly. The corresponding angular covariance map \autoref{fig:dimer_structure}\,(a2) is calculated over the dimer radii. This map shows a clear signature of a two-body breakup ($\theta_1 = \theta_2 + 180$\degree), as discussed in relation to \autoref{fig:doping_conditions}, but similarly does not allow the structure of the dimer to be determined.

\Autoref{fig:dimer_structure}\,(b1) shows an \ce{OCS^+} ion image with the alignment laser pulse, polarized along the $X$ axis, present. The expected alignment of each structure under this field is shown in \autoref{fig:potential_structures}\,(a2)-(f2). The image is still circularly symmetric, with ions being ejected in the detector ($YZ$) plane. This implies that the inter-monomer axis (IMA, the axis along which each \ce{OCS+} will recoil~\footnote{This is the $z$ axis shown in Fig. 1}) is free to rotate about the $X$ axis, such that ions are ejected in the full 360\degree~range.

The corresponding angular covariance map \autoref{fig:dimer_structure}\,(b2) shows a clear signal at $\theta_1 = \theta_2 + 180$\degree, evenly distributed over the full 360\degree~range - again consistent with a structure that is free to rotate about the $X$ axis. The observations are compatible with all structures illustrated in \autoref{fig:potential_structures}, except for the linear chain (f2), as all structures are either in, or can rotate into, a position where the IMA is in the $YZ$ plane, and can freely rotate around the $X$ axis. For the linear-chain structure the IMA would align along the $X$ axis. Thus, the \ce{OCS^+} ions would recoil perpendicular to the $YZ$ plane and end up in the center of the detector. As such the ion image does not allow us to conclude if the linear structure is present or not. To summarise, with the alignment pulse polarized along the $X$ axis none of the potential structures can be discounted.

\Autoref{fig:dimer_structure}\,(c1) shows an \ce{OCS^+} ion image with the alignment pulse polarized along the $Z$ axis. The expected alignment of each structure under this field is shown in \autoref{fig:potential_structures}\,(a3)-(f3). The image is no longer circularly symmetric. Rather the ions extend as a broad stripe along the $Y$ axis. This shows that the IMA is confined to the $XY$ plane with free rotation around the $Z$ axis. The corresponding angular covariance map \autoref{fig:dimer_structure}\,(c2) corroborates this, and shows a clear localised island at (90\degree, 270\degree) (and the equivalent at (270\degree, 90\degree) obtained by mirroring in the central diagonal). These observations are inconsistent with the T-shape structure (\autoref{fig:potential_structures}\,(e)) and the linear-chain structure (\autoref{fig:potential_structures}\,(f)). In both cases the IMA would be able to be aligned along the $Z$ axis (as drawn), thus we would expect to see a recoil at (0\degree, 180\degree) for these structures (as part of a broad stripe for the T-shape, and as a localised island for the linear-chain). We can therefore conclusively discount these structures from consideration. However, the other structures (\autoref{fig:potential_structures}\,(a3)-(d3)), are consistent with the current experimental observations. Thus, we cannot unambiguously determine which of these four structures \ce{(OCS)_2} exhibits in the droplet from the \ce{OCS^+} ion images alone.

\begin{figure}
\centering
\includegraphics[width=\columnwidth]{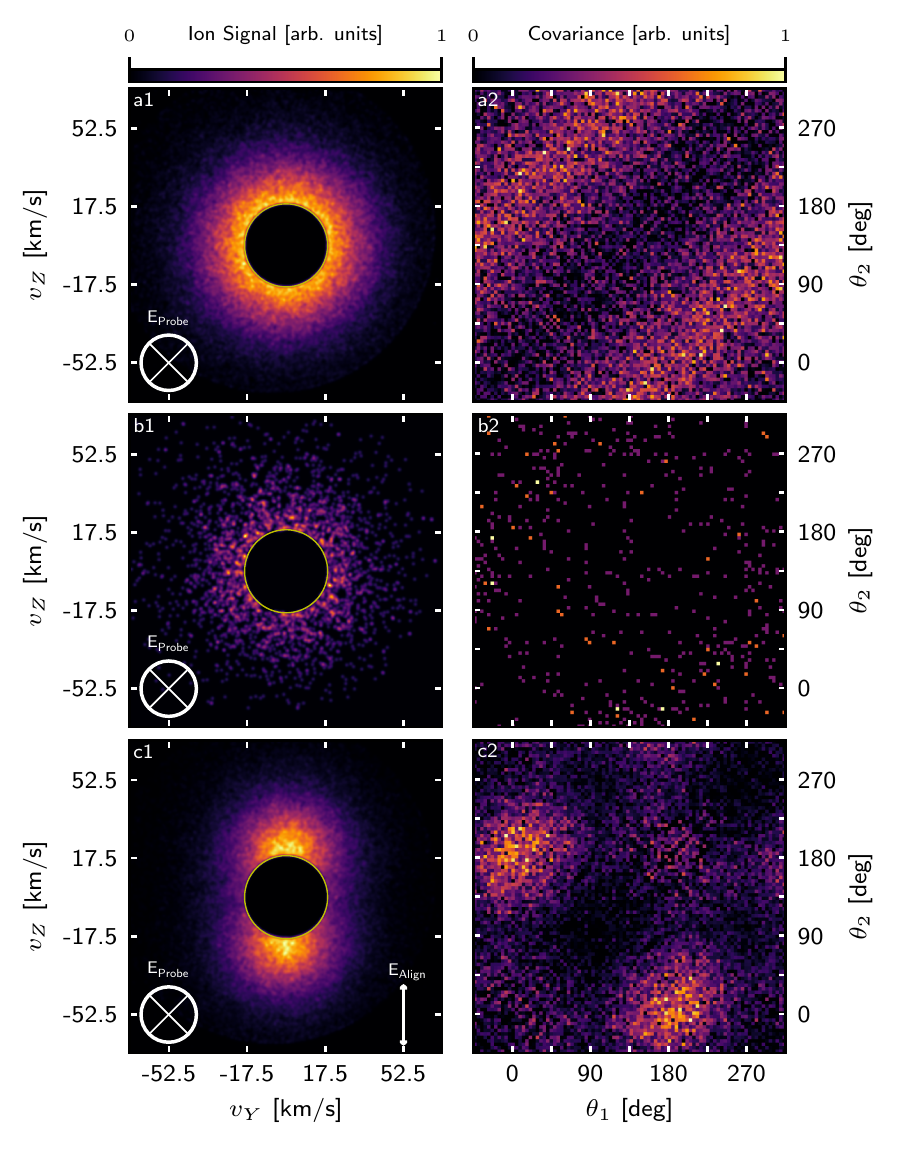}
\caption{(a1)-(c1) \ce{S^+} ion images and (a2)-(c2) angular covariance maps calculated from ions outside the yellow circle. Images (a), [(b)], (c) were taken under the dimer [monomer] doping condition. The polarization state of the probe (alignment) laser is given in the lower left (right) corner of the ion images. The center of each image is removed due to contamination by \ce{O_2^+} (see text).}\label{fig:sulphur_covariances}
\end{figure}

\Autoref{fig:sulphur_covariances}\,(a1) shows a \ce{S^+} ion image recorded with only the probe laser pulse, polarised along the $X$ axis, taken under the dimer-doping-condition. The center is removed due to the presence of singly charged molecular oxygen (\ce{O_2^+}) overlapping the \ce{S^+} channel at $m/z = \SI{32}{\dalton}$. The presence of \ce{O_2^+} results in a very intense central spot, which renders the \ce{S^+} signal difficult to see if the center is not removed. This image is also circularly symmetric, expected for a probe pulse polarized perpendicular to the detector plane. The angular covariance map \autoref{fig:sulphur_covariances}\,(a2) is calculated using all ions outside the annotated yellow circle. The observed broad stripe corresponding to $\theta_1 = \theta_2 + 180$\degree~reflects the circular symmetry of the ion image.

\Autoref{fig:sulphur_covariances}\,(b1)-(b2) show an analogous ion image and covariance map, now taken under the monomer-doping-condition. Here a marked reduction in the number of ions is observed, and almost no angular covariance signal is present. The lack of angular covariance signal for the \ce{S^+} ion is expected because Coulomb explosion of a monomer produces only one \ce{S^+} ion, i.e. there is no partner ion available for a producing a correlation event. The logical consequence of this observation is that the covariance signal observed in \autoref{fig:sulphur_covariances}\,(a2) and (c2) must arise from the Coulomb explosion of \ce{(OCS)_2}.

\Autoref{fig:sulphur_covariances}\,(c1) shows an analogous image to \autoref{fig:sulphur_covariances}\,(a1), but with the alignment pulse polarised along the $Z$ axis. Now the \ce{S^+} ions are confined along the $Z$ axis, and the angular covariance map in \autoref{fig:sulphur_covariances}\,(c2) exhibits localised island at (0\degree, 180\degree) and (180\degree, 0\degree), demonstrating that the ions recoil in opposite directions along the $Z$ axis. This `back-to-back' recoil would only be observed if the two sulfur atoms of the dimer are oriented in opposite directions along the $Z$ axis. Considering the remaining potential structures in \autoref{fig:potential_structures}\,(a3)-(d3), it is clear that only the slipped-parallel S-in (\autoref{fig:potential_structures}\,(a)) and slipped-parallel O-in (\autoref{fig:potential_structures}\,(b)) fulfill this criterion. In contrast, the slipped aligned-parallel structure has both sulfur atoms oriented in the same direction along the $Z$ axis, which would give rise to a covariance island at (0\degree, 0\degree). The cross-shaped structure would exhibit a covariance pattern with \ce{S^+} ions recoiling at around 90\degree to each other, similar to that observed for the \ce{CS_2} dimer~\cite{pickering_alignment_2018}. Thus, we can exclude these two structures, and conclude that \ce{(OCS)} dimer in \ce{He} droplets is a slipped-antiparallel structure but we cannot conclusively determine if it is the O-in or the S-in structure (\autoref{fig:potential_structures}\,(a) or (b)).

\subsection{\ce{OCS} Trimer Results}\label{sec:trimer}
\begin{figure}
\centering
\includegraphics[width=\columnwidth]{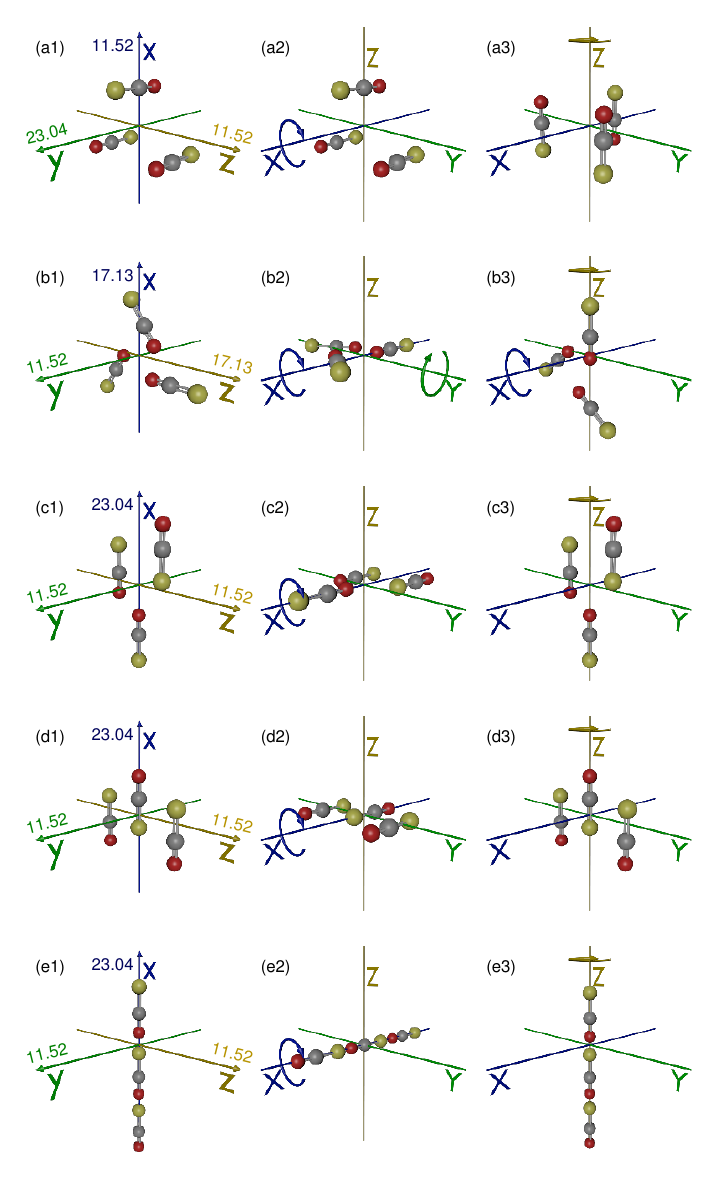}
\caption{(a1)-(e1) Sketches of potential \ce{(OCS)_3} structures in the molecular ($x,y,z$) frame, with their polarizability elements ($\alpha_x$, $\alpha_y$, $\alpha_z$) annotated on the relevant axis, in units of $\AA^3$ (see text). (a2)-(f2) [(a3)-(f3)] Sketches of how each structure would align in the laboratory ($X,Y,Z$) frame with an aligning laser field polarized along the $X$ [$Z$] axis. Curved arrows indicate axes about which rotation is possible.}\label{fig:trimer_structures}
\end{figure}

Next we show how our experimental data allow us to identify the structure of the \ce{OCS} trimer inside \ce{He} droplets. The strategy is the same as for the dimer, i.e. analyse if a selected number of potential structures are consistent with our observations. \Autoref{fig:trimer_structures}\,(a1)-(e1) show a series of possible structures for \ce{(OCS)_3}, in the molecular ($x,y,z$) frame, with polarizability components annotated on the relevant axes. As for the dimers the polarizabilities were calculated by adding the polarizability tensors of the monomers in their respective orientations. The barrel-shaped structure shown in \autoref{fig:trimer_structures}\,(a1) is known to be the ground-state structure in the gas phase; and the triangle-shaped structure (\autoref{fig:trimer_structures}\,(b1)) is observed for \ce{(CO_2)_3} in the gas phase~\cite{moazzen-ahmadi_spectroscopy_2013}. All other structures shown are not observed experimentally in the gas phase, yet are considered due to the effect of the \ce{He} environment, as discussed previously. Structures shown in \autoref{fig:trimer_structures}\,(c1) and (d1) are obtained by adding a third \ce{OCS} molecule to the ground state slipped-parallel \ce{S}-in dimer (\autoref{fig:potential_structures}\,(a1)), approaching along the molecular $x$ axis (c1) and along the molecular $z$ axis (d1), and will be referred to as the staggered-parallel and inline-parallel structures respectively. The linear chain-like structure shown in \autoref{fig:trimer_structures}\,(e1) is postulated to exist from theoretical work \cite{esrafili_abinitio_2014}, and other polar triatomic molecules in \ce{He} droplets have been shown to adopt such structures in prior spectroscopic work~\cite{nauta_nonequilibrium_1999-1}, thus we consider it here. As previously mentioned, it is germane to the discussion that it is plausible that we anneal multiple structures inside the \ce{He} droplets.

\begin{figure}
\centering
\includegraphics[width=\columnwidth]{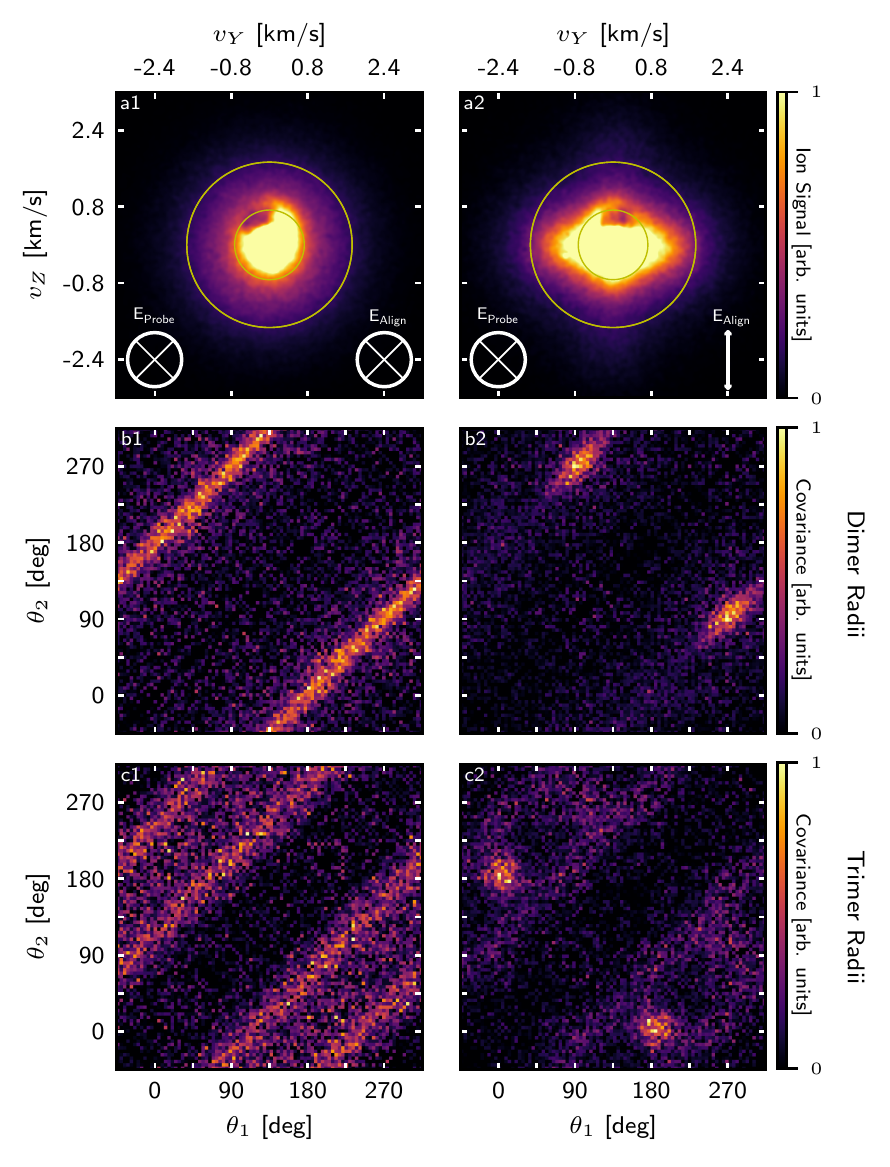}
\caption{(a1)-(a2) \ce{OCS^+} ion images and (b1)-(c2) corresponding angular covariance maps produced using data recorded under the trimer-doping-condition. Data shown in column (1) [(2)] was recorded with an aligning field polarized along the $X$ [$Z$] axis present. Covariance maps (b1)-(b2) [(c1)-(c2)] were calculated using the dimer [trimer] radii. The polarization of the probe laser is shown in the lower left corner of each ion image.}\label{fig:trimer_data}
\end{figure}

\Autoref{fig:trimer_data}\,(a1) shows an \ce{OCS^+} ion image, recorded under the trimer-doping-condition, with the alignment and probe pulse both polarised along the $X$ axis. The alignment of each structure under this field is shown in \autoref{fig:trimer_structures}\,(a2)-(e2). As mentioned in \autoref{sec:doping} there is necessarily a substantial amount of \ce{(OCS)_2} present here, and the angular covariance map calculated using the dimer radii (\autoref{fig:trimer_data}\,(b1)) shows an identical pattern to that observed in \autoref{fig:dimer_structure}\,(b2)). However, there is clearly ion signal outside the outermost annotated yellow circle (trimer radii) in \autoref{fig:trimer_data}\,(a1), which arises from Coulomb explosion of \ce{(OCS)_3} as discussed in \autoref{sec:doping}. The presence of this ion signal implies that at least one of the inter-monomer axes (IMAs) is in the $YZ$ plane, such that \ce{OCS^+} ions recoil in the $YZ$ plane as observed. Furthermore, at least one of the IMAs must be able to freely rotate around the $X$ axis to account for the circularly symmetric ion distribution. Considering the aligned sketches, this is the case for all except the linear-chain structure (\autoref{fig:trimer_structures}\,(e2)). The corresponding angular covariance map calculated using the trimer radii is shown in \autoref{fig:trimer_data}\,(c1), and shows a clear signature of a three-body breakup, with fragments being ejected at $\theta_1 = \theta_2 + 120$\degree~and $\theta_1 = \theta_2 + 240$\degree. This covariance pattern is only consistent with the barrel-shaped structure under this alignment field. Explosion of all other structures would result in at least one \ce{OCS^+} ion recoiling out of the $YZ$ plane along the $X$ axis, which would not result in the well defined emission angles in the $YZ$ plane observed in the covariance map. Specifically, \autoref{fig:trimer_structures}\,(b2)-(d2) could all result in a covariance map exhibiting a clear $\theta_1 = \theta_2 + 180$\degree\,pattern. This is not observed, however it is now pertinent to consider the kinetic energy release from various structures in more detail.

Ion events are used to calculate the trimer-radii covariance map if they occur with a speed greater than \SI{1.4}{\kilo\metre\per\second}, measured in the $YZ$ plane. If an ion event is produced with a speed greater than \SI{1.4}{\kilo\metre\per\second}, but with a substantial velocity component directed along the $X$ axis, it could have a velocity of less than \SI{1.4}{\kilo\metre\per\second} measured in the $YZ$ plane, and thus overlap with dimer radii. Therefore, the absence of covariance at $\theta_1 = \theta_2 + 180$\degree~in \autoref{fig:trimer_data}\,(c1) cannot conclusively exclude any structure that would result in this pattern, if there is a possibility that a substantial velocity component of the total ion speed is directed along the $X$ axis. For example, structures shown in \autoref{fig:trimer_structures}\,(b2) and (c2) could both produce a circular distribution of ions with a covariance signal at $\theta_1 = \theta_2 + 180$\degree, but it could overlap with the \ce{(OCS)_2} signal, thus they cannot be discounted based on this data alone. Conversely, the structure shown in \autoref{fig:trimer_structures}\,(d2) would also produce a circular distribution of ions with a covariance signal at $\theta_1 = \theta_2 + 180$\degree, except now the central \ce{OCS} in the structure will receive substantially less of the total kinetic energy (see \autoref{sec:ker}), resulting in the terminal \ce{OCS} molecules having a higher velocity in the $YZ$ plane. As a consequence of this, we would expect to see this covariance signal clearly in the trimer radii map shown in \autoref{fig:trimer_data}\,(c1). This is not the case, and thus we can conclusively discount this structure from consideration. To summarise, we have conclusively demonstrated the presence of the barrel-shaped structure (\autoref{fig:trimer_structures}\,(a2)), and conclusively discounted the presence of the structure in \autoref{fig:trimer_structures}\,(d2); but cannot rule out the other structures based on the information presented in \autoref{fig:trimer_data}\,(a1)-(c1).

\Autoref{fig:trimer_data}\,(a2) shows an \ce{OCS^+} ion image, recorded under identical conditions to \autoref{fig:trimer_data}\,(a1), but with the alignment pulse polarized along the $Z$ axis. Again there is a substantial amount of \ce{(OCS)_2} present, and the dimer radii covariance map (\autoref{fig:trimer_data}\,(b2)) is identical to \autoref{fig:dimer_structure}\,(c2). The ion signal outside of the outermost yellow circle is no longer circularly symmetric. The corresponding trimer radii angular covariance map \autoref{fig:trimer_data}\,(c2), shows a pronounced signal at (0\degree, 180\degree) and at (180\degree, 0\degree), together with some weaker structured background signal. The localised nature of these covariance islands implies that a structure in which the IMAs are aligned along the $Z$ axis is present. Only the linear-chain structure (\autoref{fig:trimer_structures}\,(a3)) is consistent with this observation, as the IMAs are aligned along the $Z$ axis under this alignment field, and the \ce{OCS^+} ions from the ends of the trimers will recoil in opposite directions along the $Z$ axis, in line with the observations. Thus we conclusively assign the (0\degree, 180\degree), (180\degree, 0\degree) covariance signals as arising from the presence of a linear-chain structure. Now considering the weaker structured background signal, this shows a similar, albeit much weaker, signal to that observed in \autoref{fig:trimer_data}\,(c1), with covariance signals at $\theta_1 = \theta_2 + 120$\degree and $\theta_1 = \theta_2 + 240$\degree. There is also a weak area of covariance signal centered around (90\degree, 270\degree). We now consider each remaining potential structure in turn with regards to these observations.

The barrel-shaped structure (\autoref{fig:trimer_structures}\,(a3)) will not produce this observed signal at $\theta_1 = \theta_2 + 120$\degree$/240$\degree, but would be expected to produce a signal at (90\degree, 270\degree), similar to the signal from \ce{(OCS)_2}. This signal will largely overlap with the signal from \ce{(OCS)_2}, but ions resulting from explosion of \ce{(OCS)_3} will gain more energy in the $YZ$ plane, but not substantially more. We therefore assign this weak signal at (90\degree, 270\degree) as arising from the barrel-shaped trimer. The trigonal-planar structure (\autoref{fig:trimer_structures}\,(b3)) could produce the signal at $\theta_1 = \theta_2 + 120/240$\degree, as it will align such that the plane containing the IMAs is in the $YZ$ plane, producing the observed covariance pattern. The structure shown in \autoref{fig:trimer_structures}\,(c3) could also give rise to this covariance signal, in an analogous way to the trigonal planar structure. We have already discounted the presence of the structure shown in \autoref{fig:trimer_structures}\,(d3) based on \autoref{fig:trimer_data}\,(c2), and it is evident that it would give rise to a broad covariance signal at $\theta_1 = \theta_2 + 180$\degree. This signal is not observed, corroborating our earlier conclusions. Therefore, we conclusively demonstrate the presence of a linear-chain structure (\autoref{fig:trimer_structures}\,(e3)) based on the data with this alignment polarization; and speculate that the weak background signal at $\theta_1 = \theta_2 + 120/240$\degree\,arises from either the trigonal planar structure (\autoref{fig:trimer_structures}\,(b3)) or the staggered-parallel structure (\autoref{fig:trimer_structures}\,(c3)). The weak background signal at (90\degree, 270\degree), arises from residual \ce{OCS^+} ions from the barrel-shaped structure.

Sketches in \autoref{fig:trimer_structures} show the individual \ce{OCS} molecules oriented in specific directions. In the case of the barrel-shaped structure (a1), this is the expected structure from the gas-phase~\cite{moazzen-ahmadi_spectroscopy_2013}, and the linear-chain structure (e1) is shown with all dipoles aligned, as was observed for \ce{HCN} inside \ce{He} droplets, and would be expected to maximise the dipole-dipole interaction; and was also calculated to maximise the chalcogen bonding~\cite{esrafili_abinitio_2014}. However, we are unable to conclusively demonstrate that this is how the \ce{OCS} molecules are oriented, as an equivalent analysis of the \ce{S^+} ion images as in \autoref{sec:dimer} is not possible under the trimer-doping-condition. This is because an \ce{S^+} ion arising from \ce{(OCS)_3} does not gain substantially more kinetic energy than an \ce{S^+} ion arising from \ce{(OCS)_2}.

To summarise, we conclusively demonstrate the observation of at least two structures of \ce{(OCS)_3} in \ce{He} droplets: the barrel-shaped structure (\autoref{fig:trimer_structures}\,(a)), previously observed in the gas phase; and the linear-chain structure (\autoref{fig:trimer_structures}\,(e)), not previously observed to our knowledge. We also speculate as to the presence of either a trigonal-planar or staggered-parallel structure, (\autoref{fig:trimer_structures}\,(b) or (c)), which has not previously been predicted for \ce{(OCS)_3}. Of these two structures, we expect that the trigonal-planar structure (\autoref{fig:trimer_structures}\,(b)) is most likely, as it has been observed for \ce{CO2} trimers previously in the gas phase~\cite{moazzen-ahmadi_spectroscopy_2013}, but the weak covariance signal means that this is only a speculation.

\subsection{Discussion}\label{sec:discussion}
The observation of the multiple structures of \ce{(OCS)_3} in \ce{He} droplets provokes considerable interest. There is a clear precedent for molecules inside \ce{He} droplets to anneal into local minima on the global potential energy surface, and there is also a precedent for multiple structures to anneal inside \ce{He} droplets~\cite{nauta_formation_2000,douberly_the_2005,nauta_nonequilibrium_1999-1}. However, complicating this is the fact that there is only one dimer structure observed in our experiments. The formation of a barrel shaped structure is easily understood by considering the addition of an extra \ce{OCS} molecule to the observed \ce{(OCS)_2} structure, and the trimer can then be formed without substantial rearrangement of the present dimer structure. A trigonal-planar structure can be formed in a similar manner. However, we conclusively showed that there is no trace of a linear dimer in \autoref{sec:dimer}. The implication of this is that the slipped-parallel dimer must rearrange upon the addition of a third \ce{OCS} molecule into the \ce{He} droplet, to enable formation of the linear-chain trimer structure. The low temperature \ce{He} environment would be expected to prohibit any such rearrangement of the dimer, however the extra energy released upon binding of a third \ce{OCS} molecule may be sufficient to allow the weakly-bound dimer to rearrange into a linear structure.

One limitation of our technique, as discussed in \autoref{sec:trimer}, is that we are unable to determine the orientation of each \ce{OCS} molecule in the barrel-shaped and linear structures. One would, however, expect that the dipole-dipole interaction and possible chalcogen bonding~\cite{esrafili_abinitio_2014} would stabilise the structure shown \autoref{fig:trimer_structures}\,(e1). Another limitation is that we are unable to determine accurate bond lengths from the kinetic energy of the fragments due to the energy loss from scattering of ions off the He droplet. It is possible that this limitation can be overcome by measurements at several different droplet sizes to quantify how much energy is lost in the scattering process.

\section{Conclusions and Outlook}\label{sec:conclusions}
In summary, the structure of dimers and trimers of \ce{OCS} molecules embedded in helium nanodroplets were identified by detecting correlations in the emission directions of the nascent \ce{OCS+} or \ce{S+} ions, following femtosecond laser-induced Coulomb explosion. The structural determination relied on confining the most polarisable axis of the dimer/trimer parallel or perpendicular to the detection plane prior to the Coulomb explosion event, using one-dimensional laser-induced adiabatic alignment. Our results show that the dimer is only formed in a slipped-parallel structure (as in the gas phase), whereas the trimer was identified in both the barrel-shaped (gas-phase) structure, and in a linear geometry. The latter, which has never been observed previously, demonstrates how \ce{He} droplets enable the formation of molecular complexes in non-ground state configurations.

For small molecules, like \ce{OCS}, the structural probing is not as complete and precise as that afforded by frequency-resolved spectroscopy. However, the interesting aspect of the Coulomb explosion method employed here is its inherent femtosecond time resolution when combined with a femtosecond pump pulse that induces dynamics, typically through photoexcitation. This opens possibilities for real-time imaging of molecular complexes undergoing structural rearrangement, including exciplex formation and bimolecular reactions.

The work presented here relies crucially on the ability to detect intact parent ions from ionization of sharply aligned molecules. The alignment occurs in the adiabatic limit implying that the parent ions created are exposed to the moderately strong ($I =$\SI{8E11}{\intensity}) alignment pulse. For \ce{OCS} reported here as well as for \ce{CS2} reported earlier~\cite{pickering_alignment_2018}, the parent ions are not affected by the presence of the alignment field. In contrast, for larger molecules, the alignment field leads to a complete fragmentation of the parent ions because the molecular cations in general absorb effectively at the \SI{800}{\nano\metre} wavelength of the alignment pulse. Recent work~\cite{chatterley_long_2018} has, however, shown that the sharp alignment at the peak of the adiabatic alignment pulse can be maintained under field-free conditions for about 10-\SI{15}{\pico\second} by rapidly turning off the pulse through spectral truncation~\cite{chatterley_communication_2018}. This renders the intact parent ions useful observables and points towards extending our method to structural determination of molecular dimers (and trimers) of much larger species than explored here. Ongoing studies in our laboratory show that this may be possible for polycyclic aromatic hydrocarbons such as tetracene.

\section{Acknowledgements\label{acks}}

We acknowledge support from the European Research Council-AdG (Project No. 320459, DropletControl) and from the European Union Horizon 2020 research and innovation program under the Marie Sklodowska-Curie Grant Agreement No. 641789 MEDEA.

%

\end{document}